\documentclass[journal]{IEEEtran}
\usepackage{amssymb, graphicx, cite, amsmath, color, array, algpseudocode, mathtools, esint, mathrsfs, eufrak, psfrag, subfigure}
\def\Gammasd{\Gamma_\text{SD}}

\def\lsd{\ell_\text{SD}}
\def\Omegasd{\Omega_\text{SD}}
\def\Gammasj{\Gamma_{\text{S},j}}
\def\gammasj{\gamma_{\text{S},j}}
\def\gammasx{\gamma_{\text{S}\mathcal{X}}}
\def\gammaxd{\gamma_{\mathcal{X}\text{D}}}
\def\Omegasj{\Omega_{\text{S},j}}
\def\Gammajd{\Gamma_{j,\text{D}}}
\def\gammajd{\gamma_{j,\text{D}}}
\def\Omegajd{\Omega_{j,\text{D}}}
\def\ljd{\ell_{j,\text{D}}}
\def\lsj{\ell_{\text{S},j}}
\def\lxd{\ell_{\mathcal{X}\text{D}}}
\def\lsx{\ell_{\text{S}\mathcal{X}}}
\def\la{\ell_\text{SR}}
\def\lb{\ell_\text{RD}}
\def\Cj{\mathscr{C}_j}
\def\Cx{\mathscr{C}_\mathcal{X}}
\def\Cja{\mathscr{C}_1}

\def\Gammasx{\Gamma_{\text{S}\mathcal{X}}}
\def\Gammaxd{\Gamma_{\mathcal{X}\text{D}}}

\def\GammaAmax{\Gamma_{\AX}^\text{max}}

\def\Omegasx{\Omega_{\text{S}\mathcal{X}}}
\def\Omegaxd{\Omega_{\mathcal{X}\text{D}}}

\def\gsr{\gamma_\text{SR}}
\def\grd{\gamma_\text{RD}}

\def\rj{\text{R}_j}
\def\rx{\text{R}_\mathcal{X}}
\def\ra{\text{R}_1}
\def\rb{\text{R}_2}
\def\r{\text{R}}
\def\kz{K_0}
\def\ps{P_\text{S}}
\def\kr{K_\text{R}}
\def\kd{K_\text{D}}
\def\prj{P_{j}}
\def\pr{P_\text{R}}

\def\nz{N_0}
\def\nr{N_\text{R}}
\def\nd{N_\text{D}}
\def\fomega{\mathrm{F}_\Omega}
\def\A{\mathcal{A}}
\def\AX{\mathcal{A}_\mathcal{\mathring X}}
\def\dx{\mathrm{d}x}
\def\ds{\mathrm{d}s}
\def\dta{\mathrm{d}\theta_1}
\def\dtb{\mathrm{d}\theta_2}

\def\thr{\gamma_\text{R}^\text{th}}
\def\thd{\gamma_\text{D}^\text{th}}
\def\Tj{T_j}

\def\Tx{T_\mathcal{X}}
\def\Ta{T_1}
\def\Tb{T_2}
\def\pave{\overline{P}_{\!j}}
\def\pavex{\overline{P}_{\!\mathcal{X}}}
\def\paveop{\overline{P}_{\!\mathcal{X}}}
\def\pavea{\overline{P}_1}
\def\paveb{\overline{P}_2}

\def\Zj{\mathcal{Q}_j} 
\def\Za{\mathcal{Q}_1} 
\def\Zb{\mathcal{Q}_2} 
\def\Zx{\mathcal{Q}_{\mathcal{X}}} 
\def\Zr{\mathcal{Q}_\text{R}}

\begin{document}
	\title{A Fair Relay Selection Scheme for a DF Cooperative Network With Spatially Random Relays}	
	\author{Masoumeh Sadeghi and Amir Masoud Rabiei	\vspace*{-2ex}
	\thanks{The authors are with the School of Electrical and Computer Engineering, College of Engineering,
	University of Tehran, P.O.Box 14395--515, Tehran, Iran.
	(e-mail:\{masoume.sadeghi, rabiei\}@ut.ac.ir).}}
\maketitle

\begin{abstract}
A new, fair relay selection scheme is proposed for a dual-hop decode-and-forward network with randomly-distributed relays. 
Most of the reported works in the literature achieve fairness at the expense of degrading the outage probability performance.  In addition, they often assume that the number and locations of the relays are known.  In contrast, the proposed scheme achieves fairness in a random field of relays without deteriorating the outage probability performance.  In this scheme, each relay maintains a countdown timer whose initial value is a function of the relay location and a tunable parameter which controls the level of fairness. The optimum value of this parameter is evaluated in an offline manner so as to achieve fairness by making the average powers consumed by the relays as close as possible.  
An exact analytical expression is derived for the average power consumed by each relay.  This expression is then used to show the superiority of the proposed scheme over opportunistic relaying and random relay selection schemes. 
\end{abstract}
	\IEEEpeerreviewmaketitle
\section{Introduction}
Cooperative communication is known to be an effective means for combating the adverse effect of multipath fading on wireless communication systems \cite{A1,A2}. In a   
cooperative system, idle users (referred to as relays in the sequel) serve as virtual antennas for the source and destination nodes allowing them to achieve spatial diversity \cite{g1}.
%
%
\par Cooperative communication was first introduced in \cite{azari1} and then further investigated in \cite{azari2} and more recently in \cite{A2}. 
Several cooperation protocols have been proposed in the literature among which amplify-and-forward (AF) and decode-and-forward (DF) have received much attention owing to their simplicity and effectiveness \cite{A2, A3}.   
In order to take better advantage of relaying, signals received from the source and relays have to be properly combined at destination.  
Opportunistic relaying \cite{g2}, 
is a simple yet effective combining scheme in which only a single relay participates in cooperation.  This relay should be among the set of relays that can correctly decode the source signal.  In addition, it should have the largest relay-destination signal-to-noise power ratio (SNR).
%
An important but less investigated issue in relay networks is to design a \textit{fair} relaying strategy, i.e., a strategy in which the average powers consumed by the relays are approximately the same.  Despite its many advantages, opportunistic relaying suffers from lack of fairness among relays, i.e., a relay with \textit{slightly} better average channel gain than others is always chosen for cooperation and, hence, its power drains much faster than those of other relays. 
%
%
As a result, it is crucial to design a fair relay assignment strategy in which the relays participating in cooperation consume approximately the same amount of power \cite{b19, g, b}.
%
\par In \cite{b19}, a fair relay selection technique has been proposed for an AF relaying network which attempts to equally divide the total consumed power among  relays. This technique, however, has a larger outage probability than that of the opportunistic relaying except when the received SNR is large, or when the number of relays is small.  Hence, in this technique the quality of service (QoS) may be compromised to establish fairness among the relays.
In contrast to \cite{b19}, in \cite{g} a relay selection scheme (known as outage priority based proportional fair scheduling) has been proposed which gives a higher priority to outage probability than fairness.  This scheme improves fairness among relays without degrading the outage probability performance compared to opportunistic relaying. Again, this improvement is significant only when the SNR is large. 
%
%
In \cite{b}, a fair power allocation scheme has been proposed under outage probability constraint. In this scheme, each relay has a threshold value which is used to determine whether it can participate in cooperation or not.  The optimum values of the thresholds are obtained by solving an optimization problem that minimizes the total power consumption under a set of constraints imposed by the QoS and fairness requirements.
\par Although there is a large body of research focused on fair cooperative networks, in most of them the total number of relays and their positions are assumed to be fixed.  However, in many practical scenarios the relays are mobile.  Therefore, it is reasonable to assume that the relays are distributed randomly in their deployment region. 
In references \cite{R1,R2}, the outage performance of AF and DF relaying protocols are investigated for the case where the relays are distributed as a homogeneous two-dimensional (2D) Poisson point process (PPP) with constant density. An exact statistical analysis has been conducted for the distance between a reference node and its communication best neighbor in a Poisson field of nodes in \cite{aydin2015}. 
In \cite{azari}, a probabilistic relay assignment strategy has been proposed which decreases the total power consumed by the relays through reducing the number of relay deployments.
\par In this paper, we propose a fair relay selection strategy for a two-hop DF relaying network with randomly distributed relays. To the best of authors' knowledge, only few works reported in the literature consider fairness among relays in a random field of relays.
Our relay selection scheme aims to improve fairness among relays while achieving the same average outage probability as does the opportunistic relaying. { To this end, we first form a set of relays that can successfully forward the source signal to destination.  Each relay in this set has a countdown timer whose initial value is determined in a way that a desired level of fairness among relays is achieved. The initial value of each relay's timer is a function of the relay's location and a tunable parameter, referred to as $\beta$, that controls the level of fairness among relays. 
We derive an exact analytical expression for the average power consumed by each relay and use this result to find the optimum value of $\beta$.  We also use this expression to demonstrate the superiority of our proposed scheme over opportunistic relaying and random relay selection schemes. 
among relays.}
\par The remainder of this paper is organized as follows.  In Section \ref{system model}, the system model is introduced.  Our fair relay selection scheme is proposed in Section \ref{part1}.  In Section \ref{part2}, we derive an exact solution for the average power consumed by an arbitrary relay in the network.  Numerical results are presented in Section \ref{part3}.  Concluding remarks are given in Section \ref{part4}. 
\begin{figure}[!t]
    	\centering
   \includegraphics[width=0.9\linewidth]{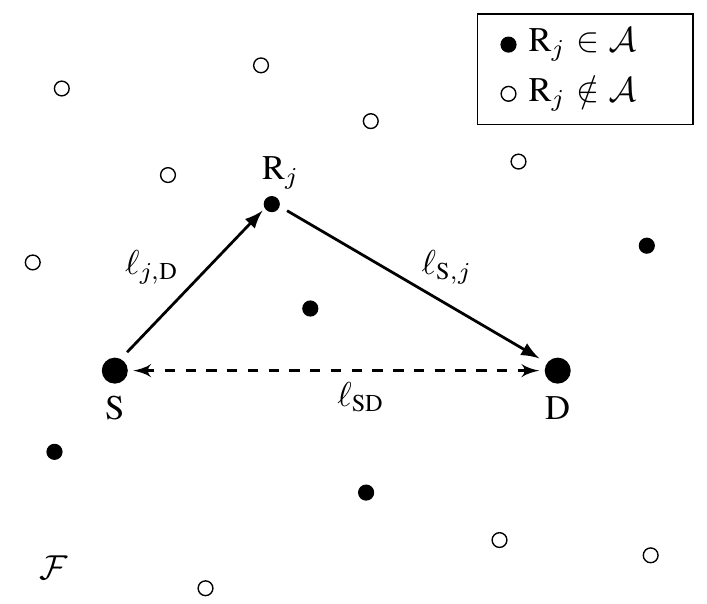}
	\caption{A typical dual-hop DF relay network.}
	\label{fig1}
\end{figure}
	\section{System Model}
	\label{system model}
	A typical dual-hop DF relay network is shown in Fig. \ref{fig1}. 
	This network consists of a source node, S, a destination node, D, at distance $\ell_\text{SD}$ of the source node and a number of relays.  As mentioned earlier, we assume that the relays constitute a 2D homogeneous PPP with constant intensity $\lambda$.  {For simplicity of our analysis, we assume the relay deployment region, $\mathcal{F}$, is arbitrarily large.  Clearly, when a relay is far from the source and destination nodes, the attenuation due to path loss becomes quite significant. As a result, such relay is quite unlikely to take part in cooperation \cite{R1}.} 
	\par Suppose that the transmission channels are subject to path loss and small scale fading.  Then, the instantaneous SNR of the S--D link is given by \cite{R1}
	\begin{subequations}
		\begin{equation}
			\Gammasd=\frac{\kz\,\ps}{\nz\lsd ^{\alpha}}\Omegasd
		\end{equation}		
	\end{subequations}
where $\ps$ is the source power, $\alpha\in[2,7]$ is the path loss exponent and $\kz$ is a constant that is determined by the system parameters (e.g., antenna gain), $\nz$ is the noise power at destination, and $\Omegasd$ is the fading power in the S--D channel with $\mathbb{E}\{\Omegasd \}=1 $. Denoting by $\rj$ the $j$th relay in the system, the instantaneous SNRs of  the S--R$_j$ and R$_j$--D links are 
	\begin{subequations}
		\begin{align} \label{gsj}
			\Gammasj&=\gammasj\,\Omegasj\\ \label{gjd}
			\Gammajd&=\gammajd\,\Omegajd
		\end{align}
	\end{subequations}
	where $\Omegasj$ and $\Omegajd$ are the fading powers of S--$\rj$ and $\rj$--D links, respectively.  We assume  $\Omegasj$ and $\Omegajd$  are independent unit-mean random variables with the same cumulative distribution function (CDF) $\fomega(\cdot)$. Also in \eqref{gsj} and \eqref{gjd}
	\begin{subequations}
		\begin{align}
			\gammasj&=\frac{\kr \,\ps}{\nr\,\lsj^{\alpha}}\\
			\gammajd&=\frac{\kd \,\pr}{\nd\,\ljd^{\alpha}}
		\end{align}		
	\end{subequations}%
where $\pr$ is the transmitted power of $\rj$ when it is chosen for cooperation, $\lsj $ and $ \ljd $ are S--$\rj$ and $\rj$--D  distances, respectively.  In addition, $\kr$ and $\kd$ are constants that depend on the system parameters, and $\nr$ and $\nd$ are the noise powers at relays and destination, respectively. 
\par In this paper, we consider a dual-hop DF relaying system in which a relay $\rj$ is qualified for cooperation provided that
	\begin{enumerate}
		\item $\rj$ can correctly decode the received signal from the source, i.e., $\Gammasj\geq\thr$
		\item destination node can correctly decode the signal that is forwarded by $\rj$ , i.e., $\Gammajd\geq\thd$  
	\end{enumerate}
where $\thr$ and $\thd$ are given thresholds that are properly set so as to meet the system requirements.
	As a result, the set of all such relays is defined as
	\begin{equation}\label{mya}
		\A\triangleq\big\{\rj\big|\Gammasj\geq\thr,~\Gammajd\geq\thd\big\}.
	\end{equation}
	If $\A$ is an empty set, no cooperation takes place, i.e., an outage occurs in the network.  When $\A$ is not empty, each relay in this set is assigned a timer and the relay whose timer expires first is selected for cooperation. All other relays in $\A$ are then requested to back-off \cite{g2}. 
	 The timer assignment strategy depends on the relaying scheme that is used.  For example, in opportunistic relaying the relay which provides the largest SNR at destination is chosen for cooperation.  Therefore, in opportunistic relaying, the $j$th relay's timer should be inversely proportional to $\Gammajd$. 
	Clearly, in opportunistic relaying, relays with good channel conditions in both S--R and R--D links are more likely to be selected for cooperation.  This can put the burden of cooperation on some specific relays and result in poor fairness among relays.  
	\section{Fair Relay Selection}
	\label{part1}
In this section, we assume that fairness is achieved when the average powers consumed by the relays are approximately the same.  Then, we propose a relay selection scheme that aims to make the average consumed powers as close as possible. 
\par	Suppose that $\Cj$ denotes the event that R$_j$ is chosen for cooperation. Then, we define $\prj$ as
	\begin{equation}
		\prj\triangleq\Bigg\{\hspace*{-1ex}%
		\begin{array}{ll}%
			\!\pr, &\Cj \text{ occurs}\\
			0,   &\text{otherwise}.
		\end{array}
	\end{equation}
	Using the total probability theorem we obtain the average of $\prj$ as
	\begin{align}
		\pave &\triangleq \mathbb{E}\{\prj\}\notag =\mathbb{E}\{\prj|\Cj\}\Pr\{\Cj\}+\mathbb{E}\{\prj|\Cj^\text{c}\} \Pr\{\Cj^\text{c}\}\notag\\
		&= \Pr\{\Cj\}\pr\label{qq1}
	\end{align}
	where $\Cj^\text{c}$ is the complement of $\Cj$.
	Now, we use the fact that $\Pr\{\Cj|\rj\notin \A\}=0$ along with the total probability theorem to obtain
	\begin{align}
		\Pr\{\Cj\} &= \Pr\{\Cj|\rj\in \A\}\Pr\{\rj\in \A\}\notag \\&\hspace*{15ex}+\Pr\{\Cj|\rj\notin \A\}\Pr\{\rj\notin \A\} \notag\\ 
		&= \Pr\{\Cj|\rj\in \A\}\Pr\{\rj\in \A\} \label{qq2}.
	\end{align}
	Substituting for $\Pr\{\Cj\}$ from \eqref{qq2} into \eqref{qq1} and denoting $\Pr\{\rj\in \A\}$ by $\Zj$, one arrives at
	\begin{align}
		\pave&=\Pr\{\Cj|\rj\in \A\} \,\Zj\,\pr \label{qq5}.
	\end{align}
	Note that $\Zj$ can be evaluated using \eqref{mya} along with the fact that $\Omegasj$ is independent of $\Omegajd$ as 
	\begin{subequations}
		\begin{align}
			\Zj&=\Pr\{\rj\in \A\}=\Pr\big\{\Gammasj \geq\thr,~\Gammajd \geq\thd\big\}\notag\\
			&=\Pr\left\{\Omegasj\geq\frac{\thr}{\gammasj}\right\}\,\Pr\left\{\Omegajd\geq\frac{\thd}{\gammajd}\right\}\notag\\
			&=\mathbb{Y}\left(\frac{\thr}{\gamma_\text{SR}}\Big(\frac{\lsj}{\lsd}\Big)^{\alpha},\frac{\thd}{\gamma_\text{RD}}\Big(\frac{\ljd}{\lsd}\Big)^{\alpha}\right)\label{qq37}
		\end{align}
where
\begin{align}
			\mathbb{Y}(u,v)&\triangleq(1-\fomega(u))(1-\fomega(v))\label{qq21}\\
			\gamma_\text{SR}&\triangleq\frac{\kr\,\ps}{\nr\lsd^{\alpha}}\\
			\gamma_\text{RD}&\triangleq\frac{\kd\,\pr}{\nd\lsd^{\alpha}}
\end{align}
	\end{subequations}
	and, recall, $\fomega(\cdot)$ is the CDF of $\Omegasj$ and $\Omegajd$. 
	\par 
	In the next subsection we consider a simple relay network with only two relays and propose a timer assignment scheme for this network.  Then, we consider a random network in which the number and the locations of the relays are random and extend our proposed timer assignment strategy to this case.
	\subsection{Two-Relay Network} \label{two-relay}
	Consider a simple relay network with only two relays, namely, $\ra$ and $\rb$.  These relays are assumed to be randomly placed in the S--D plane.  As seen in \eqref{qq37}, $\Zj$ is a deterministic function of $\lsj$ and $\ljd$. Thus, without loss of generality, we can assume that, on average, $\ra$ has better channel conditions than $\rb$ which, in turn, implies that $\Za>\Zb$, i.e., R$_1$ is more likely to be in $\A$ than R$_2$. Hence, when both $\ra$ and $\rb$ are in $\A$, a fair relay selection strategy should select R$_2$ more frequently than R$_1$.
	In consequence, the timer assigned to R$_1$ should have a larger initial value.  A simple solution is to set the initial time of R$_j$'s timer, i.e., $\Tj$, directly proportional to $\Zj$, i.e.,
	\begin{equation}
		\Tj=\mathcal{C}\, \Zj,\quad j=1,2\label{qq84}
	\end{equation}
	where $\mathcal{C}$ is a constant used to adjust the timer duration within a given range. Observe that for the case where $\Za$ is only slightly greater than $\Zb$, eq. \eqref{qq84} implies that $\Ta>\Tb$ and, thus, when both relays are in $\A$, $\rb$ is always selected for cooperation.  This causes a poor fairness between $\ra$ and $\rb$. 
	In order to achieve fairness between $\ra$ and $\rb$ in this case, a random term has to be considered in $\Tj$'s so as to prevent the all-time selection of $\rb$. Hence, eq. \eqref{qq84} is written as 
	\begin{equation}
		\Tj=\mathcal{D}_j\,\Zj,\quad j=1,2\label{qq85}
	\end{equation}
	where $\{\mathcal{D}_j\}$'s are independent random variables uniformly distributed between $0$ and $\mathcal{C}$, i.e., $\mathcal{D}_j \sim \mathcal{U}(0,\mathcal{C})$. It is clear from \eqref{qq85} that $\{\Tj\}$'s are independent random variables and that $\Tj\sim \mathcal{U}(0,\mathcal{C}\Zj)$.
	In order to find $\pavea$ we first note that 
	\begin{align} \notag
		&\Pr\{\Cja|\ra\in A\}\notag\\ &= \Pr\{\Cja|\ra\in A \text{ and } \rb\in A\}\Pr\{\rb\in\A|\ra\in\A\}\notag\\&\quad\qquad+\Pr\{\Cja|\ra\in \A \text{ and } \rb\notin \A\}\Pr\{\rb\notin\A|\ra\in\A\} \notag \\& = \Pr\{\Ta<\Tb\} \,\Pr\{\rb\in\A\}+1 \times\Pr\{\rb\notin\A\} \label{pr1}
	\end{align} 
	where \eqref{pr1} follows from the fact that the events $\{\ra\in\A\}$ and $\{\rb\in\A\}$ are independent.
	Substituting from \eqref{pr1} into \eqref{qq5} for $j=1$, one arrives at
	\begin{align}
		\pavea&= \left(1-\Zb \Pr\{\Ta\ge \Tb\}\right)\Za\,\pr\label{eq2}\\
		\shortintertext{and similarly}
		\paveb&=\left(1-\Za \Pr\{\Tb\ge \Ta\}\right)\Zb\,\pr.\label{eq3}
	\end{align}
	Recalling that $\Ta$ and $\Tb$ are independent and uniformly distributed over $[0,\mathcal{C}\Za]$ and $[0,\mathcal{C}\Zb]$, respectively, one can readily show that
	\begin{equation}
		\Pr\{\Ta\ge\Tb\}=1-\Pr\{\Tb\ge\Ta\}=1-\frac{\Zb}{2\Za}\label{eq4}.
	\end{equation}
	Therefore, \eqref{eq2} and \eqref{eq3} can be written as
	\begin{align}
		\pavea&= \pr\,\left(\Za-\Zb\left(\Za-\frac{\Zb}{2}\right)\right)\label{eq6}\\
		\paveb&= \pr\,\Zb\left(1-\frac{\Zb}{2}\right)\label{eq7}
	\end{align}
	respectively. Observe from \eqref{eq6} and \eqref{eq7} that for $\Za\approx\Zb$, the average powers of $\ra$ and $\rb$ are almost the same.  However, when $\Za$ and $\Zb$ are not close to each other, $\pavea$ and $\paveb$ can be quite different. In order to address this issue, we add a new parameter, i.e., $\beta>0$, to the definition of $\Tj$ as
	\begin{equation}
		\Tj=\mathcal{D}_j\,\Zj^\beta,\quad j=1,2.\label{qq86}
	\end{equation}
	Choosing the timers according to \eqref{qq86}, one obtains 
	\begin{align}
		\pavea&=\pr\, \Za\left(1-\Zb+\frac{\Zb}{2}\Big(\frac{\Zb}{\Za}\Big)^{\!\beta}\right)\label{eq66}\\
		\paveb&=\pr\, \Zb\left(1-\,\frac{\Za}{2}\Big(\frac{\Zb}{\Za}\Big)^{\!\beta}\right).\label{eq77}
	\end{align}
	Clearly, for the case where $\Za>\Zb$, increasing $\beta$ increases $\pavea$ but decreases $\paveb$ and vice versa.  As a result, we can adjust $\beta$ so as to 
	make $\pavea$ and $\paveb$ close to each other. It is important to note that, in general, $\pavea$ and $\paveb$ cannot become arbitrarily close, i.e., we can not find $\beta$ so that $\big|\pavea-\paveb\big|=0$. To clarify this point, we observe from \eqref{eq66} and \eqref{eq77} that for $\Za>\Zb$ and $\beta\rightarrow\infty$, $\pavea\rightarrow\pr\Za(1-\Zb)$ and $\paveb\rightarrow\pr\Zb$. Now, if $\Za$ and $\Zb$ are such that $\Za>\Za(1-\Zb)>\Zb$,\footnote{This occurs when $\rb$ is always chosen for cooperation whenever it is in $\A$.  In consequence, $\ra$ takes part in cooperation only when $\rb\notin\A$.} then $\pavea\ne\paveb$ even for $\beta\rightarrow\infty$.  In other words, in this case we always have $\big|\pavea-\paveb\big|\ge\big(\Za-\Zb-\Za\Zb\big)\pr$.

	\subsection{Poisson Relay Network}
	\par We now turn our attention to the case where the potential relays distributed as a 2D homogeneous PPP with density $\lambda$.  
	Again, our objective is to make the $\{\pave\}$'s as close as possible. 
	Recall from \eqref{qq5} that this objective is met when $\Pr\{\Cj|\rj\in A\} \,\Zj$ is approximately the same for all relays.  As a result, to have a fair relay selection, relays with larger $\Zj$ should be selected less frequently.  This means that the timer duration, $\Tj$, should be proportional to $\Zj$. On the other hand, as discussed in Subsection \ref{two-relay}, when $\{\Zj\}$'s are not close to each other, a parameter $\beta$ should be considered in $\Tj$ (as given by \eqref{qq86}) to provide a higher level of fairness among relays.
	Thus, with $\{\Tj\}$'s defined as in eq. \eqref{qq86}, the problem is to find $\beta$ so as to achieve a reasonable level of fairness among relays.%
\footnote{Note that a special case, referred to as \textit{random relay selection}, emerges when $\beta$ is set to zero, i.e., all relays in $\A$ are equally likely to be selected for cooperation.  Expectedly, random relay selection is not a fair selection scheme as the cooperation probability for each relay in this scheme will be proportional to the probability that the relay is within $\A$. Thus, relays that are more likely to be in $\A$ have, on average, a larger cooperation probability.}
\par An effective means to determine $\beta$ is to minimize the maximum unfairness in the network. 
	This will reduce the burden of cooperation from the relays with good channel conditions and distribute it fairly among other relays.
	Suppose that $\mathcal{X}$ is an arbitrary point in the S--D plane whose distances from the source and destination nodes are given by $\ell_{\text{S}\mathcal{X}}$ and $\ell_{\mathcal{X}\text{D}}$, respectively.  From eqs. \eqref{qq5} and \eqref{qq37} it is clear that the average power consumed by a relay located at $\mathcal{X}$, referred to as $\pavex$ in the following, depends on the location of $\mathcal{X}$.  
	Therefore, we define the maximum unfairness as 
	\begin{equation} \label{unfairness}
		\max_{\lsx,\lxd}\{\pavex\}-\min_{\lsx,\lxd}\{\pavex\}
	\end{equation}
	where the maximum and minimum in \eqref{unfairness} are obtained over all locations that $\mathcal{X}$ can be placed in S--D plane.  Observe that in a random network, relays that are very far from source and destination nodes have very little chance to be selected for cooperation meaning that their average consumed power is approximately equal to zero. Hence, $\min\limits_{\ell_{\text{S}\mathcal{X}},\;\ell_{\mathcal{X}\text{D}}}\{\pavex\}$ is approximately equal to zero and we can rewrite the maximum unfairness of the network as  
	$\max\limits_{\ell_{\text{S}\mathcal{X}},\;\ell_{\mathcal{X}\text{D}}}\{\pavex\}$. 
	%
	%
	%
	%
	As a result, our problem reduces to find an optimum value for $\beta$ as
	\begin{equation} \label{betaopt}
		\beta^\text{opt}=\arg\,\min_{\beta}\Big\{\max_{\ell_{\text{S}\mathcal{X}},\;\ell_{\mathcal{X}\text{D}}}\{\pavex\}\Big\}.
	\end{equation}
Therefore, in order to obtain $\beta^\text{opt}$, we need to evaluate $\pavex$ as a function of $\beta$, $\lsx$ and $\lxd$. 
%
	\section{Evaluation of $\pavex$} 
	\label{part2}
In this section we derive an analytical expression for the average power consumed by $\rx$, i.e., a relay located at $\mathcal{X}$.  In Subsections \ref{proposed} and \ref{opportunistic}, we derive $\pavex$ for our proposed relay selection scheme and opportunistic relaying, respectively.
	\subsection{Proposed Relay Selection Scheme} \label{proposed}       
\begin{figure}[!t]
		\centering
\includegraphics[width=.9\linewidth]{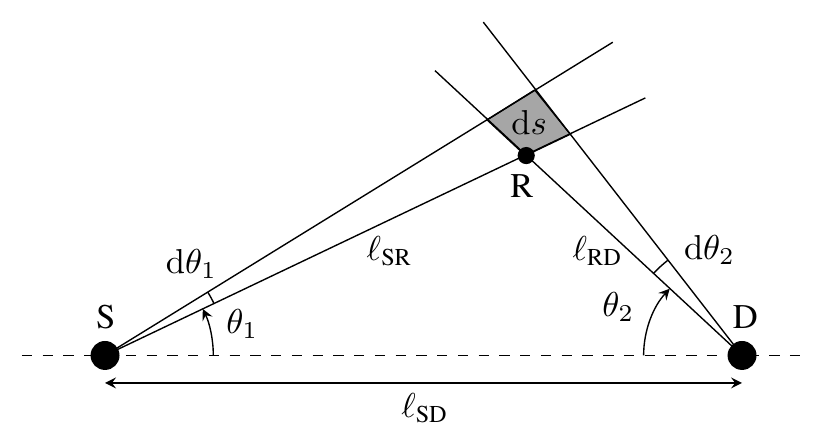}
		\caption{Illustration of the biangular coordinate system.}
		\label{Biangular}
	\end{figure}

We assume, without loss of generality, that $\mathcal{C}$ in \eqref{qq84} equals unity. We also assume that the timer allocated to $\rx$ has a duration of $\Tx$, which, recall from \eqref{qq86}, is uniformly distributed over $\big[0,\,\Zx^\beta\big]$ and $\Zx$ is defined in \eqref{qq37} with $\lsj$ and $\ljd$ replaced by $\lsx$ and $\lxd$, respectively.  Denoting by $\Cx$ the event that $\rx$ is chosen for cooperation, and using eq. \eqref{qq5} along with the total probability theorem, one can write 
	\begin{align}
		\pavex &= \pr \Zx \int_{0}^{\Zx^\beta} f_{\Tx}(x\,|\,\rx\in\A)\notag\\&\qquad\qquad\qquad\qquad\times\Pr\{\Cx|\rx\in \A,\Tx=x\}\,\dx \notag \\
		&= \frac{\pr \Zx}{\Zx^\beta}\int_{0}^{\Zx^\beta}\Pr\{\Cx|\rx\in \A,\Tx=x\}\;\dx. \label{qq6}
	\end{align}
Therefore, we need to obtain $\Pr\{\Cx|\rx\in \A,\Tx=x\}$. From the properties of the PPPs, it is known that all other relays in the network except for $\rx$ are again distributed according to a two-dimensional PPP with density $\lambda$. Assume now  that $\AX$ is defined as
	\begin{equation}\label{ax}
		\AX\triangleq\big\{\rj\ne \rx\big|\Gammasj\geq\thr,~\Gammajd\geq\thd\big\}.
	\end{equation}
	Then, we can make use of marking theorem \cite[p. 55]{kingman} along with \eqref{qq37} to show that the relays in the set $\AX$ are distributed as a 2D nonhomogeneous PPP with mean measure
	\begin{align}
		\mu(\ds)&=\Pr\{\r\in\AX\}\,\lambda\,\ds=\Zr\,\lambda\,\ds\label{density}
	\end{align}
	where $\ds$ is the surface element, $\r$ is an arbitrary relay in the S--D plane whose distances from the source and destination nodes are given by $\la$ and $\lb$, respectively, and $\Zr$ equals
	\begin{align}
		\Zr=\mathbb{Y}\left(\frac{\thr}{\gamma_\text{SR}}\Big(\frac{\la}{\lsd}\Big)^{\alpha},\frac{\thd}{\gamma_\text{RD}}\Big(\frac{\lb}{\lsd}\Big)^{\alpha}\right)\label{zr}.
	\end{align} 
	Suppose now that $\rx$ is in $\A$ and $\Tx$ is equal to $x$. Then $\rx$ is selected for cooperation if either $\AX$ is an empty set or the timers of all relays in $\AX$ are greater than $x$. As a result, we can write
	\begin{equation}
		\Pr\{\Cx\,|\,\rx\!\in\!A,\Tx=x\}
		=\mathbb{E}\bigg\{\prod_{\rj\in\AX}\!\!\Pr\big\{\Tj> x|\AX\big\}\bigg\}\label{eq10}
	\end{equation}
	where the expectation in \eqref{eq10} is over all realizations of $\AX$. Hence, if we choose $\{T_j\}$'s according to eq. \eqref{qq86} along with \eqref{qq37}, we obtain
	\begin{equation}
		\Pr\{\Tj\geq x|\A\} =\left\{ \begin{array}{lr}
			\!\!1-\frac{x}{\Zj^\beta},& x<\Zj^\beta\\
			\!\!0, & x\ge \Zj^\beta
		\end{array}\right.
=\Big[1-x\,\Zj^{-\beta}\Big]^+ \label{aa}
	\end{equation}
	where $[x]^+\triangleq \max\{0,x\}$. Substituting for $\Pr\{\Tj\geq x|\A\}$ from \eqref{aa} into \eqref{eq10} and using \cite[eq. (3.35)]{kingman}, one arrives at
	\begin{align}
		\Pr\{\Cx\,|\,\rx\in \A,\Tx=x\}&\notag\\ &\hspace*{-12ex}=\exp\bigg[-\int_{\mathcal{F}}\left(1-\Big[1-x\,\Zr^{-\beta}\Big]^+\right)\,\mu(\ds)\bigg]\notag\\
		&\hspace*{-12ex}=\exp\bigg[-\int_{\mathcal{F}}\Big[x\,\Zr^{-\beta}\Big]^{\le1}\,\mu(\ds)\bigg] \notag \\
		&\hspace*{-12ex}=\exp\bigg[-\lambda\,\int_{\mathcal{F}}\Big[x\,\Zr^{-\beta}\Big]^{\le1}\,\Zr\,\ds\bigg] \label{aa1}
	\end{align}
	where $[x]^{\le1}\triangleq \min\{1,x\}$ and, recall from \eqref{zr}, $\Zr$ is a function of $\la$ and $\lb$. In order to evaluate the integral in \eqref{aa1} we use a \textit{biangular coordinate system} \cite{A26}.  As shown in Fig. \ref{Biangular}, in this coordinate system, given two poles S and D with distance $\lsd$, any point R (except for those located on the axis of abscissas) can be uniquely represented by a pair of angles $(\theta_1,\theta_2)$ where $-\pi\le\{\theta_1,\theta_2\}\le\pi$, $\theta_1\cdot\theta_2\ge0$ and $|\theta_1+\theta_2|\le\pi$. Now, using the law of sines in $\overset{\triangle}{\text{SRD}}$, we obtain \cite{R1}
	\begin{align}
		\la&=\lsd\frac{\sin(\theta_2)}{\sin(\theta_1+\theta_2)}\label{qq25}\\
		\lb&=\lsd\frac{\sin(\theta_1)}{\sin(\theta_1+\theta_2)}\label{qq26}\\ \label{qq28}
		\mathrm{d}s&=\lsd^2\Big|\frac{\sin(\theta_1)\,\sin(\theta_2)}{\sin^3(\theta_1+\theta_2)}\Big|\dta\dtb
	\end{align}
	where $\la$ and $\lb$  are S--R and R--D distances, respectively. 
	Considering the symmetry of our model about the S--D crossing line, the integral in \eqref{aa1} can be evaluated only for the upper half of $\mathcal{F}$, which is characterized as $0\le\theta_1\le\pi$ and $0\le\theta_2\le\pi-\theta_1$. 
	Using \eqref{zr} along with \eqref{qq25} through \eqref{qq28} in \eqref{aa1} one arrives at
	\begin{subequations}
	\begin{equation}
		\Pr\{\Cx|\rx\in \A,\Tx=x\}=\exp\!\bigg[\!-\lambda\,\lsd^2\mathbb{G}\Big(\frac{\gamma_{\text{SR}}}{\thr},\frac{\gamma_{\text{RD}}}{\thd},x\Big)\bigg]\label{qq30}
	\end{equation}
	where $\mathbb{G}(\cdot,\cdot,\cdot)$ is given by
	\begin{flalign} \notag
		&\mathbb{G}(u,v,x)\triangleq \\&2\!\int_{0}^{\pi}\int_{0}^{\pi-\theta_1}
		\!\!\bigg[x\,\mathbb{Y}\Big(\frac{1}{u}\frac{\sin^{\alpha}(\theta_2)}{\sin^{\alpha}(\theta_1\!+\!\theta_2)}\,,\frac{1}{v}\frac{\sin^{\alpha}(\theta_2)}{\sin^{\alpha}(\theta_1\!+\!\theta_2)}\Big)^{\!\!\raisebox{-2pt}{$\scriptstyle-\beta$}}\bigg]^{\le1}\!\!\times \notag\\
		&\mathbb{Y}\Big(\frac{1}{u}\frac{\sin^{\alpha}(\theta_2)}{\sin^{\alpha}(\theta_1\!+\!\theta_2)}\,,\frac{1}{v}\frac{\sin^{\alpha}(\theta_2)}{\sin^{\alpha}(\theta_1\!+\!\theta_2)}\Big)
		\frac{\sin(\theta_1)\sin(\theta_2)}{\sin^3(\theta_1\!+\!\theta_2)}\,\dtb \dta.\label{qq29}
	\end{flalign}
	\end{subequations}
	Finally, substituting from \eqref{qq30} into \eqref{qq6} gives
	\begin{equation}
		\pavex=\dfrac{\pr}{\Zx^{\beta-1}}\int_{0}^{\Zx^{\beta}}\exp\!\bigg[\!-\bar{n}\,\mathbb{G}\Big(\frac{\gamma_\text{SR}}{\thr},\frac{\gamma_\text{RD}}{\thd},x\Big)\bigg]\dx\label{qq23}
	\end{equation}
	where we define $\bar{n}\triangleq\lambda\,\lsd^2$ as the normalized relay density.	Note, importantly, that \eqref{qq23} implies that $\pavex$ depends on $\ell_{\text{S}\mathcal{X}}$ and $\ell_{\mathcal{X}\text{D}}$ only through $\Zx$.  Therefore, eq. \eqref{betaopt} can be further simplified to
	\begin{equation} \label{betaopt1}
		\beta^\text{opt}=\arg\,\min_{\beta}\Big\{\max_{0<\Zx\le1}\{\pavex\}\Big\}.
	\end{equation}
	{
The optimum value of $\beta$ is then obtained by substituting for $\pavex$ from \eqref{qq23} into \eqref{betaopt1}. Note, importantly, that $\pavex$ does not depend on the locations of the relays as it is a function of $\bar n$, $\overline\gamma_\text{SR}$ and $\overline\gamma_\text{RD}$.  Hence, the mobility of nodes has no effect on $\beta^\text{opt}$, and this parameter can be evaluated in an offline manner and used by a simple table lookup. 
The main difficulty with solving eq. \eqref{betaopt1} is that $\beta$ can vary from $0$ to $\infty$. However, a thorough numerical inspection of \eqref{betaopt1} reveals that in a practical scenario when $\beta$ becomes greater than $10$, the relay with the smallest $\Zj$ is almost always selected for cooperation. Therefore, the search interval for $\beta$ can be reduced to $[0, 10]$, which can significantly reduce the complexity of solving equation \eqref{betaopt1}.   
	}
\subsection{Opportunistic Relaying}  \label{opportunistic}
In an opportunistic relaying scheme, the relay in $\A$ with maximum SNR at destination is selected for cooperation. 
Recalling that $\Cx$ denotes the event that $\rx$ is chosen for cooperation, and that $\rx$'s transmission power equals $\pr$, one can write
\begin{align}
	\paveop&=\pr\,\Pr\{\Cx\}\notag\\
	&=\pr\,\Pr\Big\{\rx\in\mathcal{A},\Gammaxd>\GammaAmax\Big\}\\
    &=\pr\,\Pr\Big\{\Gammasx>\thr,\Gammaxd>\thr,\Gammaxd>\GammaAmax\Big\}\label{bb}
\end{align}
where $\GammaAmax\triangleq\max\{\Gammajd|\rj\in\AX\}$ and $\AX$ was defined in \eqref{ax}. Also, $\Gammasx$  and $\Gammaxd$ are defined in \eqref{gsj} and \eqref{gjd}, respectively, with $j$ replaced by $\mathcal{X}$. 
Using the fact that the fading powers of all channels  are independent random variables, we obtain
\begin{align}
		\paveop&= \pr\,\Pr\bigg\{\Omegasx	>\frac{\thr}{\gammasx},\Omegaxd>\frac{\thr}{\gammaxd},  \Omegaxd>\frac{\GammaAmax}{\gammaxd}\bigg\}\notag\\
		&=\pr\Pr\!\bigg\{\!\Omegasx\!>\!\frac{\thr}{\gammasx}\bigg\}\Pr\!\bigg\{\!\Omegaxd\!>\!\max\bigg[\frac{\thr}{\gammaxd},\frac{\GammaAmax}{\gammaxd}\bigg]\bigg\}\notag\\
		&=\pr \bigg[1-\mathrm{F}_{\Omega}\Big(\frac{\thr}{\gammasx}\Big)\bigg]
\Pr\!\bigg\{\!\Omegaxd\!>\!\max\bigg[\frac{\thr}{\gammaxd},\frac{\GammaAmax}{\gammaxd}\bigg]\bigg\}.\label{bb3}
\end{align}
Now one can use the total probability theorem to write 
\begin{align}
\Pr\!\bigg\{&\!\Omegaxd\!>\!\max\bigg[\frac{\thr}{\gammaxd},\frac{\GammaAmax}{\gammaxd}\bigg]\bigg\}\notag\\&=\int_{0}^{\infty}\Pr \bigg\{x>\frac{\thr}{\gammaxd},x>\frac{\GammaAmax}{\gammaxd}\bigg\}f_{\Omega}(x)\,\dx\notag\\
&=\int_{\frac{\thr}{\gammaxd}}^{\infty}\Pr\bigg\{x>\frac{\GammaAmax}{\gammaxd}\bigg\}f_{\Omega}(x)\dx\notag\\
&=\int_{\frac{\thr}{\gammaxd}}^{\infty}\mathbb{E}\Bigg\{\prod\limits_{\rj\in\AX}\Pr\bigg\{x>\frac{\Gammajd}{\gammaxd}\bigg\}\Bigg\}f_{\Omega}(x)\dx\label{AX1}\\
&=\int_{\frac{\thr}{\gammaxd}}^{\infty}\mathbb{E}\Bigg\{\prod\limits_{\rj\in\AX}\Pr\bigg\{x>\Omegajd\Big(\frac{\lsx}{\ljd}\Big)^\alpha\bigg\}\Bigg\}f_{\Omega}(x)\dx \label{bb4}
\end{align}	
where expectations in \eqref{AX1} and \eqref{bb4} are over all realizations of $\AX$. 
Recalling from \eqref{density} that the relays in $\AX$ constitute a nonhomogeneous PPP with mean measure $\mu(\ds)=\Zr\lambda\ds$ 
and using \cite[eq. (3.35)]{kingman}, one arrives at
	\begin{multline}
\mathbb{E}\Bigg\{\prod\limits_{\rj\in\AX}\Pr\bigg\{x>\Omegajd\Big(\frac{\lsx}{\ljd}\Big)^\alpha\bigg\}\Bigg\}\\=\exp\Bigg[-\lambda\int_{\mathcal{F}}\bigg[1-\mathrm{F}_{\Omega}\bigg(x\,\Big(\frac{\lb}{\lxd}\Big)^\alpha\bigg)\bigg]\,\Zr\ds\Bigg].\label{bb6}
	\end{multline}	
In order to evaluate the integral in \eqref{bb6} we use the biangular coordinate system again along with eqs. \eqref{zr} and \eqref{qq25}--\eqref{qq28} to obtain
	\begin{subequations}
	\begin{multline}
\mathbb{E}\Bigg\{\prod\limits_{\rj\in\AX}\Pr\bigg\{x>\Omegajd\big(\frac{\lsx}{\ljd}\big)^\alpha\bigg\}\Bigg\}\\=\exp\!\Bigg[\!-n\,\mathbb{H}\bigg(\frac{\gamma_{\text{SR}}}{\thr},\frac{\gamma_{\text{RD}}}{\thd},x\Big(\frac{\lsd}{\lxd}\Big)^\alpha\bigg)\Bigg]\label{bb8}
	\end{multline}
		where $\mathbb{H}(\cdot,\cdot,\cdot)$ is given by
	\begin{multline}
	\mathbb{H}(u,v,x)\triangleq\ 2\,\int_{0}^{\pi}\int_{0}^{\pi-\theta_1}\bigg[1-\mathrm{F}\bigg(x\,\Big(\frac{\sin(\theta_1)}{\sin(\theta_1+\theta_2)}\Big)^\alpha\bigg)\bigg]\times\\
\mathbb{Y}\bigg(\frac{1}{u}\frac{\sin^{\alpha}(\theta_2)}{\sin^{\alpha}(\theta_1\!+\!\theta_2)}\,,\frac{1}{v}\frac{\sin^{\alpha}(\theta_2)}{\sin^{\alpha}(\theta_1\!+\!\theta_2)}\bigg)
	\frac{\sin(\theta_1)\sin(\theta_2)}{\sin^3(\theta_1\!+\!\theta_2)}\,\dtb \dta.\label{bb9}
	\end{multline}
	\end{subequations}
Now, using \eqref{bb8}, \eqref{bb4} and \eqref{bb3} we get
\begin{multline}
			\paveop=\pr\bigg[1-\mathrm{F}_\Omega\Big(\frac{\thr}{\gammasx}\Big)\bigg]
			\\\times\!\int_{\frac{\thr}{\gammaxd}}^{\infty}\exp\!\bigg[\!-n\,\mathbb{H}\bigg(\frac{\gamma_{\text{SR}}}{\thr},\frac{\gamma_{\text{RD}}}{\thd},x\Big(\frac{\lsd}{\lxd}\Big)^\alpha\bigg)\bigg]f_{\Omega}(x)\dx.\label{bb7}
\end{multline}
\section{Numerical Results}
\label{part3}
In this section, we use computer simulation to confirm the validity of our analytical results and to demonstrate the effectiveness of our relay selection scheme. Simulation results are obtained using Mont-Carlo method for ten million independent realizations of the network. The path loss exponent, $\alpha$, is assumed to be $4$ and a Rayleigh fading model considered for all transmission channels, i.e.,
\begin{equation}
		\mathrm{F}_{\Omega}(x)=1-e^{-x},\quad x\ge0.
\end{equation} 
\begin{figure}
		\centering
		\includegraphics[width=.975\linewidth]{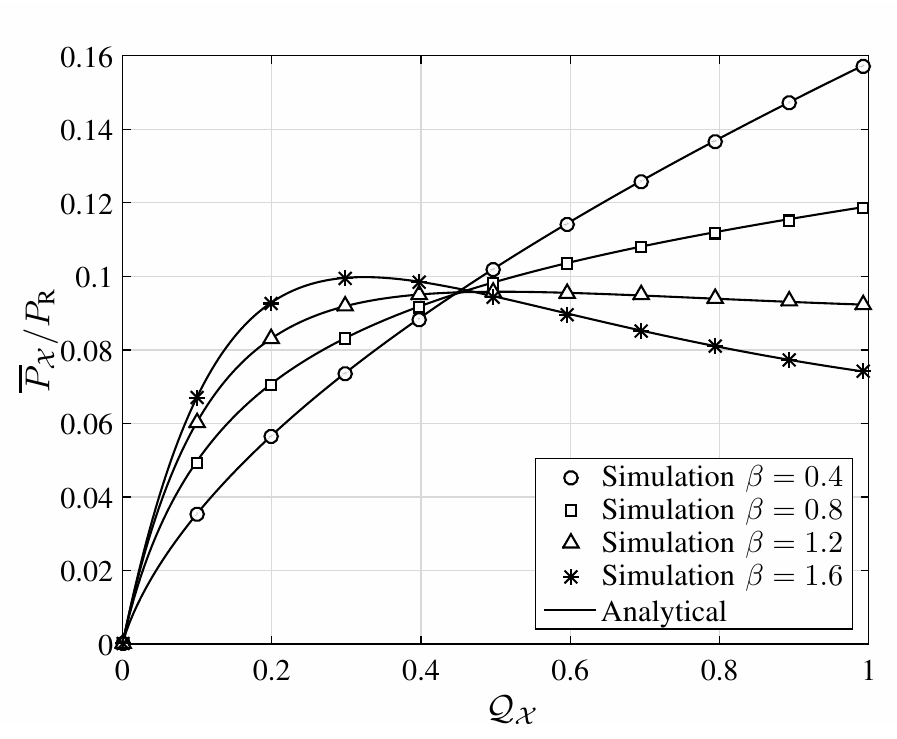}
		\caption{$\pavex/\pr$ as a function of $\Zx$ for $\gamma_\text{SR}=\gamma_\text{RD}=20$ dB, $\bar n=2$ and $\thr=\thd=5\,\text{dB}$.}
		\label{sym1}
\end{figure}
Fig. \ref{sym1} shows $\pavex/\pr$ as a function of $\Zx$ for our proposed relay selection scheme with $\beta=0.4,\,0.8,\,1.2$ and $1.6$, $\gamma_\text{SR}=\gamma_\text{RD}=20\,\text{dB}$, $\bar n=2$ and $\thr=\thd=5\,\text{dB}$.  Recall that all relays consume the same amount of power, i.e., $\pr$, when they are selected for cooperation.  Hence, $\pavex/\pr$ is, indeed, the average probability that a relay is selected for cooperation.
As seen in Fig. \ref{sym1}, our analytical results are in complete agreement with the simulation results for all the examined values of $\beta$.  
Note that a high degree of fairness is achieved when $\pavex/\pr$ remains approximately unchanged for all values of $\Zx$.  Fig. \ref{sym1} also indicates that different levels of fairness can be achieved by changing the value of $\beta$.  For example, for $\beta=1.2$ our relay selection scheme provides a high degree of fairness whereas for $\beta=0.4$ it is quite unfair.  Fig. \ref{sym1} also highlights the important fact that perfect fairness cannot be achieved among relays.  This is because relays that are far from the source and destination are very unlikely to be in $\A$ and, thus, their corresponding $\pavex/\pr$ will be very close to $0$.
	\begin{figure}
		\centering
		\includegraphics[width=.975\linewidth]{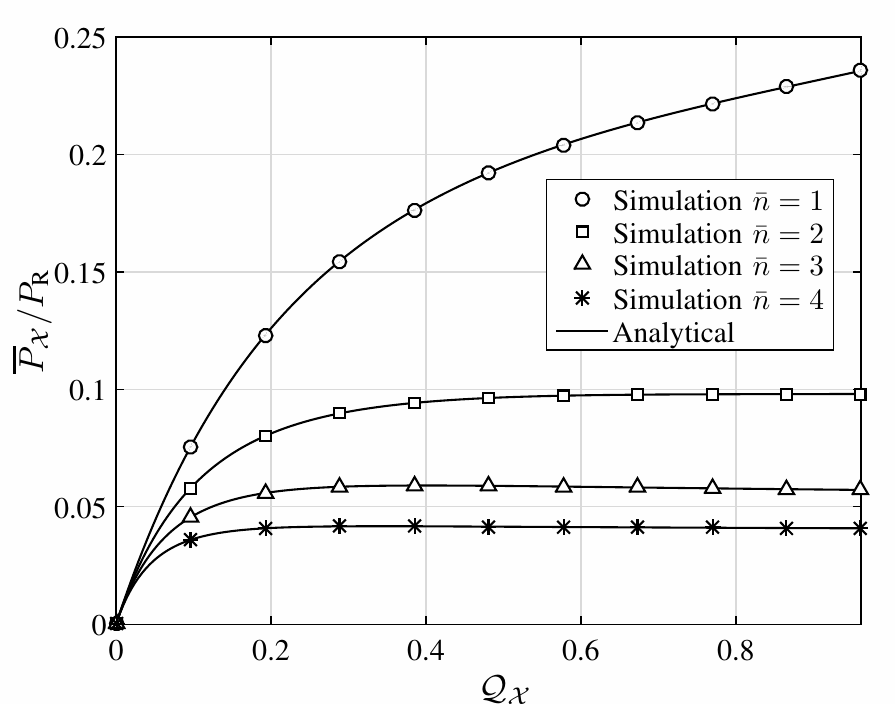}
		\caption{The cooperation probability of $\rx$ as a function of $\Zx$, $\gamma_\text{SR}=\gamma_\text{RD}=10\,\text{dB}$, $\thr=\thd=5\,\text{dB}$ and selected values of $\bar n$.}
		\label{sym2}
	\end{figure}
\par Fig. \ref{sym2} illustrates the cooperation probability of $\rx$, i.e., $\pavex/\pr$, as a function of $\Zx$ for our proposed relay selection scheme with $\bar n=1,\,2,\,3$ and $4$, $\gamma_\text{SR}=\gamma_\text{RD}=10$ dB and $\thr=\thd=5$ dB.  For each $\bar n$, the corresponding optimum value of $\beta$ has been used.  Again, our analytical results match the simulation results quite well for the examined values of $\bar n$.  As seen in this figure, a $\bar n$ increases from $1$ to $4$, the range of $\Zx$ values for which a reasonable fairness among relays can be achieved also increases. 
Note also that for a given $\Zx$ value, increasing $\bar n$ reduces the cooperation probability of each relay.  This is due to the fact that as $\bar n$ increases the number of potential relays in $\A$ also increases.  As a result, the probability that a relay in $\A$ is chosen for cooperation decreases.
\begin{figure}[!t]
		\centering
		\includegraphics[width=.975\linewidth]{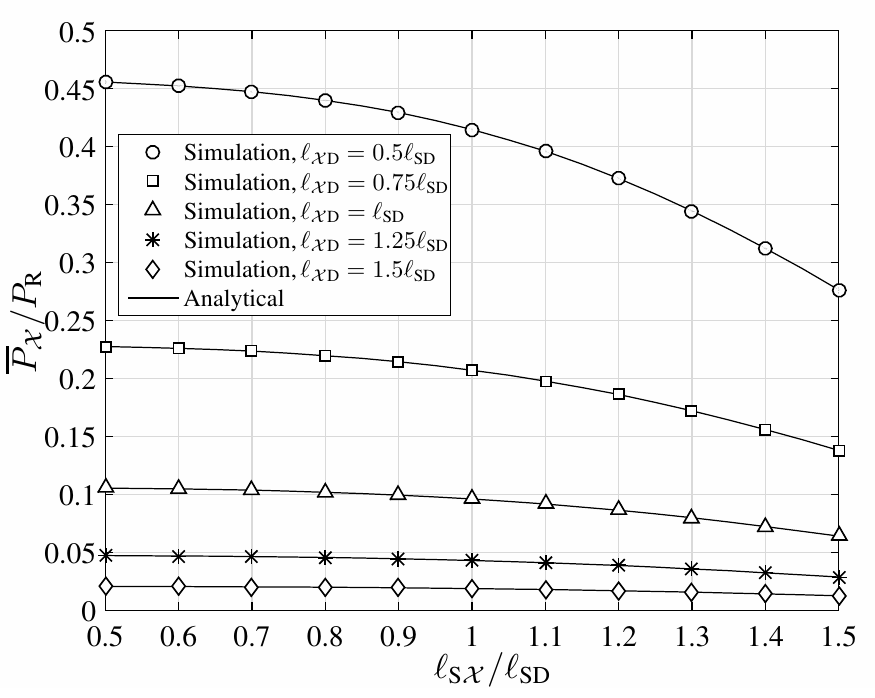}
		\caption{$\pavex/\pr$ as a function of $\lsx/\lsd$ for $\gamma_\text{SR}=\gamma_\text{RD}=15$ dB, $\bar n=2$ and $\thr=\thd=5\,\text{dB}$ in an opportunistic cooperative network.}
		\label{sym7}
\end{figure}
\par Fig. \ref{sym7} shows $\pavex/\pr$ as a function of $\lsx/\lsd$ for $\lxd/\lsd=0.5,\,0.75,\,1,\,1.25$ and $1.5$ in a system with opportunistic relaying, $\gamma_\text{SR}=\gamma_\text{RD}=15\,\text{dB}$, $\bar n=2$ and $\thr=\thd=5\,\text{dB}$.
As can be seen in Fig. \ref{sym7}, our analytical results are in complete agreement with the simulation results for all the examined scenarios.  
Expectedly, when $\lxd$ is fixed, relays that are located near source are more likely to be selected for cooperation.  Similarly, when $\lsx$ is fixed, relays that are close to destination are more likely to be selected for cooperation.  This is due to the fact that the closer is $\rx$ to source (destination), the smaller is the path loss in S--$\rx$ ($\rx$--D) channel.
	\begin{figure}[!t]
		\centering
		\includegraphics[width=0.975\linewidth]{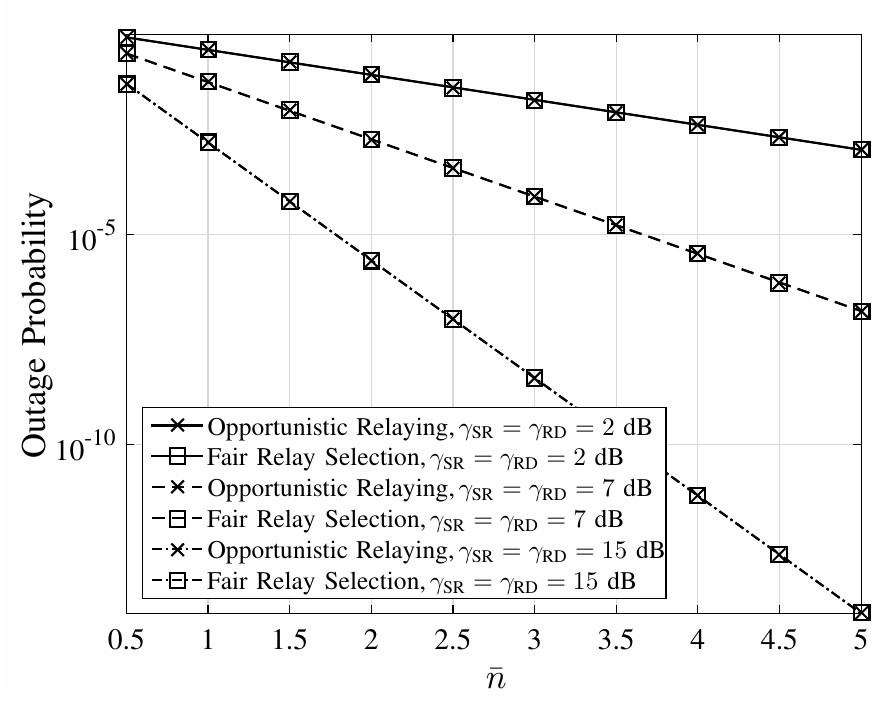}
		\caption{Outage probability of the opportunistic relaying and the proposed scheme as a function of $\bar n$ for $\gsr=\grd=2$, $7$ and $15$ dB and $\thr=\thd=3$ dB.}
		\label{sym4}
	\end{figure}
	
\begin{figure}[!t]
		\centering
		\subfigure[]{\label{op}\includegraphics[width=0.9\linewidth, draft=false, clip]{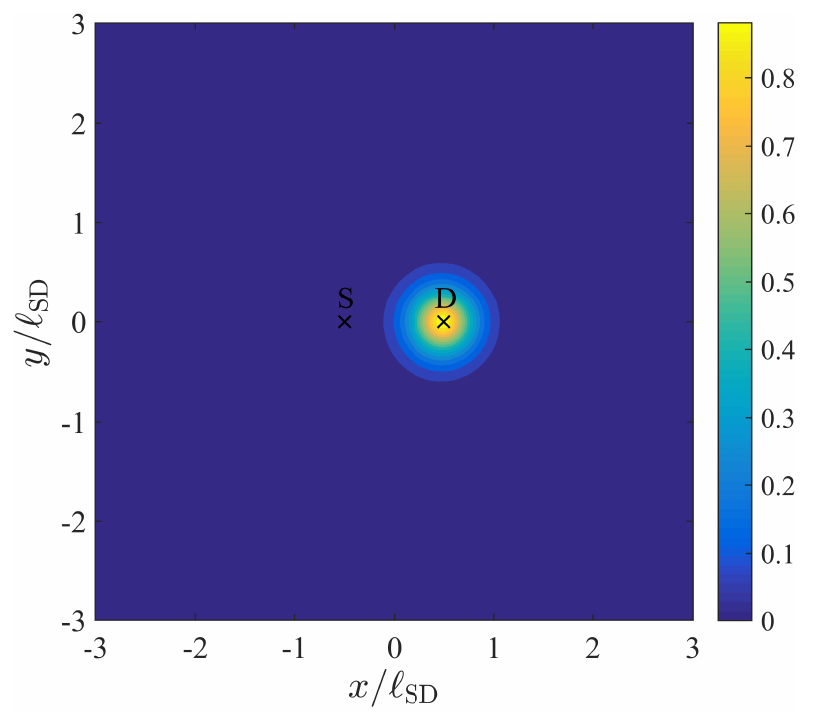}}\\
		\subfigure[]{\label{random}\includegraphics[width=0.9\linewidth, draft=false, clip]{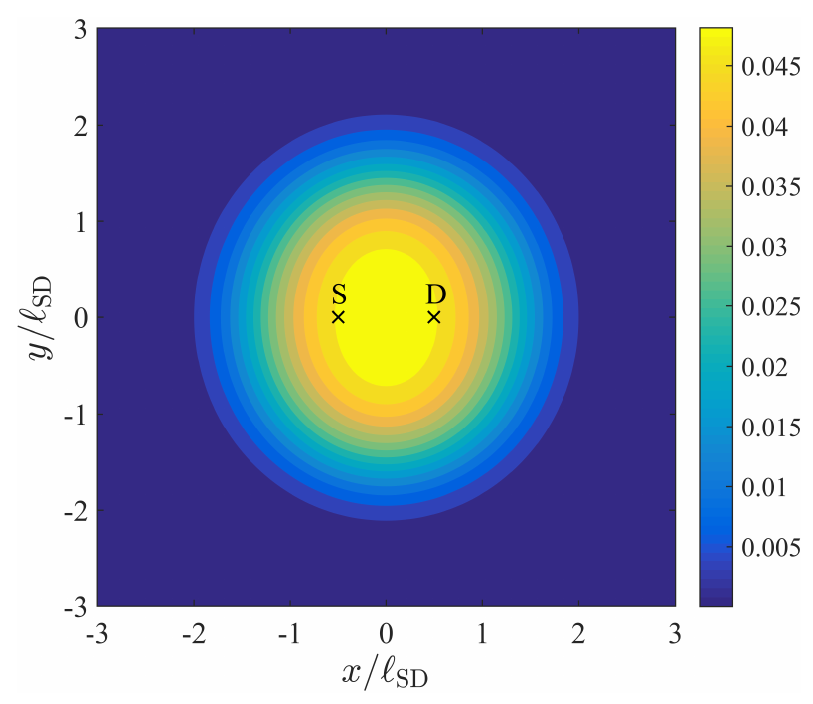}}\\
		\subfigure[]{\label{mine}\includegraphics[width=0.9\linewidth, draft=false, clip]{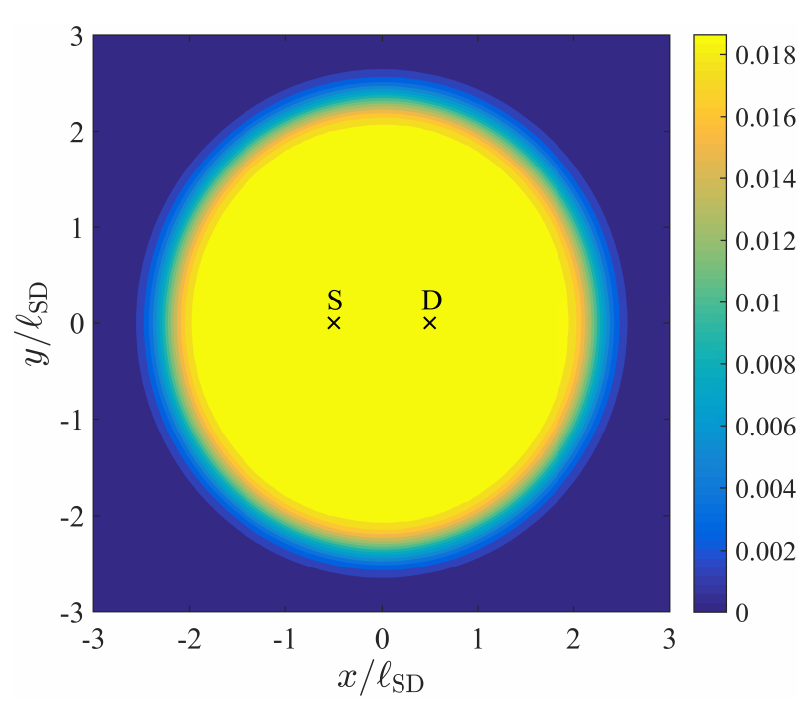}}\\		
		\caption{The cooperation probability as a contour plot in the $x$--$y$ plane for $\gsr=\grd=15$ dB, $\thr=\thd=3$ dB, $\bar n=3$, and (a) opportunistic relaying, (b) random relay selection, and (c) proposed relay selection.}
		\label{sym3}
\end{figure}
\par The outage probability of the proposed relay selection scheme and the opportunistic relaying as a function of $\bar{n}$ are illustrated in Fig. \ref{sym4}.  The results are shown for $\thr=\thd=3$ dB and $\gamma_\text{SR}=\gamma_\text{RD}=2,\,7$ and $15$ dB.  Clearly, the outage probability of both schemes are the same for all three scenarios.  To explain this, we note that in both schemes an outage occurs when the set $\A$ is empty, i.e., there is no relay that can correctly decode the source signal, or if there is such relay, the SNR of the corresponding relay-destination link is less than $\thd$.
	Also, it can be seen that in both schemes, increasing $\bar n$ reduces the outage probability. 
\par Figs. \ref{op} through \ref{mine} illustrates the cooperation probability as a contour plot in the $x$--$y$ plane for opportunistic relaying, random relay selection and the proposed relay selection schemes. It is assumed that $\gamma_\text{SR}=\gamma_\text{RD}=15\,\text{dB}$, $\thr=\thd=3\,\text{dB}$ and $\bar n=3$.  As seen in Fig. \ref{op}, opportunistic relaying provides the worst fairness level among the relays.  Indeed, in opportunistic relaying scheme the cooperation probability for a limited group of relays located around destination is greater than $0.8$, while other relays are rarely selected for cooperation, i.e., their corresponding cooperation probabilities are approximately equal to zero.  
As shown in Figs. \ref{op} and \ref{random}, random relay selection achieves a higher level of fairness compared to opportunistic relaying.  To explain this, we observe that in random relay selection scheme, relays that are located between source and destination are more likely to be in $\A$ and, thus, selected for cooperation with approximately the same probability.  Finally, Fig. \ref{mine} highlights the fact that our proposed scheme can provide the highest level of fairness among relays in comparison with opportunistic relaying and random relay selection schemes.  
%
\section{Conclusion}
\label{part4}
In this paper, a new fair relay selection scheme was proposed for a dual-hop DF relaying network with randomly-distributed relays in a general fading environment. In our proposed scheme, we first created a list of relays that could successfully forward the source signal to destination.  Then, we assigned a timer to each relay in the list and evaluated the initial values of the timers in a way that the average powers consumed by the relays become approximately the same.
An exact analytical formula for the average power consumed by each relay was derived for opportunistic relaying and our proposed relay selection scheme. 
It was shown that this formula can be used to tune a parameter in our scheme that controls the level of fairness. Simulation results showed that our proposed scheme can improve the level of fairness among relays without deteriorating the outage probability performance.  It was also observed that as the density of relays increases, our proposed scheme can provide fairness among a wider range of relays compared to opportunistic relaying and random relay selection schemes. 

\bibliographystyle{IEEEtran}
\bibliography{sadeghi}
\end{document}